\documentclass[prl,aps,preprint,showpacs,floatfix]{revtex4}
\usepackage{graphicx}

\begin{document}
\title{Quantum entropy dynamics for chaotic systems beyond the 
classical limit} 
\author{Arnaldo Gammal$^{(a)}$ and Arjendu K. Pattanayak$^{(b)}$} 
\date{\today}
\affiliation{(a)Instituto de F\'isica, Universidade de S\~{a}o Paulo,
CEP 05315-970 C.P. 66318, S\~ao Paulo-Brasil\\
(b) Department of Physics and Astronomy, Carleton College, Northfield, 
Minnesota 55057}

\begin{abstract}
The entropy production rate for an open quantum system with a classically 
chaotic limit has been previously argued to be independent of $\hbar$ and 
$D$, the parameter denoting coupling to the environment, and to be equal to 
the sum of generalized Lyapunov exponents, with these results applying
in the near-classical regime. We present results for a specific system 
going well beyond earlier work, considering how these dynamics are altered 
for the Duffing problem by changing $\hbar,D$ and show that the entropy 
dynamics have a transition from classical to quantum behavior that scales, 
at least for a finite time, as a function of $\hbar^2/D$. 

\end{abstract}
\maketitle

Consider a quantum system with a nonlinear classical limit: 
Non-classical effects depend on the size of Planck's constant 
$\hbar$ compared to the characteristic action. Further, the 
system-environment interaction as measured through some parameter $D$, 
is crucial\cite{zurek-rmp}. The dynamics of the classical limit of the 
problem are important\cite{ClassicalLimit} particularly through the 
classical Lyapunov exponents $\lambda$. It has recently 
been found that the quantum entanglement rate for chaotic systems shows 
a valuable speed-up\cite{entanglement-chaos} but this is to be balanced 
against the observed enhanced decoherence effects for chaotic systems 
in the classical limit\cite{zp,akp,MP}. However, there is 
increased stability against fidelity decay deep in the quantum parameter 
regime\cite{Prosen}, leading to the proposal to `chaoticize' quantum 
computation\cite{Prosen-qc}.  This complex multi-parameter
quantum-classical transition is fundamental, poorly understood, and also 
valuable in understanding the behavior of quantum devices.

A recent analysis\cite{akp-03}, summarized below, suggested that headway 
could be made in characterizing the full range of behavior by considering 
composite parameters and scaling. That is, the quantum-classical difference 
as measured by some quantity $QC_d(\hbar,D,\lambda)$ should be the simpler 
function $QC_d'(\zeta)$ of a single composite parameter 
$\zeta =\hbar^\alpha D^\beta\lambda^\gamma$. Evidence has 
begun to accumulate\cite{scaling-new} supporting this perspective. These
come mostly from studying the effect of changing $D,\hbar$ on 
time-independent (usually from $t\to\infty$) measures $QC_d$. 
The change with $\lambda$ is harder to study since the classical 
phase-space changes along with $\lambda$. 
A different but related issue is the non-equilibrium statistical 
mechanics of a nonlinear quantum system as measured through the 
system's entropy dynamics. A powerful result of broad interest is 
that the entropy production rate for an open quantum system with 
a classically chaotic limit is independent of $\hbar$ and $D$ 
and is equal to the sum of generalized Lyapunov 
exponents\cite{zp,akp,MP}. However, this has been verified only 
in the classical limit, and despite the considerable interest in 
this, there are few useful results away from this limit. 

In this Letter, we start with the argument that quantum-classical distance 
can be measured sensibly with a quantum system's linear entropy.  We 
then study the entropy dynamics for the chaotic Duffing oscillator as 
a function of $\hbar, D$ to obtain several novel results that considerably 
extend results on entropy decay as well as generalize the scaling results. 
Specifically, the Lyapunov exponent dependence is shown to be valid 
only for a small parameter range and for times. We look, more 
usefully, at the time-dependent entropy itself which unexpectedly
shows scaling with a single parameter $\zeta_0 = \hbar^2/D$, thus 
generalizing previous results from time-independent 
measures\cite{akp-03,scaling-new}. That is, behavior from widely 
varied $\hbar,D$ collapse onto curves that depend only on $\zeta_0$, 
which we explain on the basis of an expansion in $\zeta_0$, as well as 
direct comparison of dynamics. This enables the characterization of
entropy dynamics over a much wider range of parameters and times 
than previously attempted. We show dynamical regimes which we term 
(I) classical, (II) semi-classical and (III) quantum, with a smooth
transtion between these regimes with increasing $\zeta_0$.

We begin with the Master equation for the evolution of a quantum Wigner 
quasi-probability $\rho_W$ under Hamiltonian flow with potential $V(q)$ 
while coupled to an external environment~\cite{zp}:
\begin{eqnarray}
{\partial \rho_W\over\partial t}
&=& L_c + L_{q} + T \\
&=&\{H,\rho_W\} \nonumber \\
&+&\sum_{n \geq 1}\frac{\hbar^{2n}(-1)^n}{2^{2n} (2n +1)!}
\frac{\partial^{2n+1} V(q)}{\partial q^{2n+1}}\;
\frac{\partial^{2n+1} \rho_W}{\partial p^{2n+1}}\nonumber \\
&+& 2\gamma\partial_p(p\rho_W) + D \partial_p^2\rho_W
\label{wigner}
\end{eqnarray}
The first term, the Poisson bracket $L_c$, generates the classical 
evolution for $\rho_W$. The terms in $\hbar$ are the quantal 
`correction' terms (denoted $L_{q}$). The environmental coupling ($T$) 
is modelled by the diffusive $D$ term and the dissipative $\gamma$ term. 
We assume, as typical, short time-scales or high temperatures such 
that the $\gamma$ term is negligible. A $QC_d$ can then be considered 
by propagating the same initial condition with $L_q+L_c+T$, compared to using 
only $L_c$, or more appropriately using $L_c +T$ from above. 

If $QC_d$ is the difference between the expectation values of an 
observable, it becomes strongly dependent on the observable. For 
example, even when the centroids of a quantum and classical 
distribution are behaving identically, differences exist in higher-order 
moments. Further, measures such as the time when the $QC_d$ hits a 
pre-defined value introduce subjectivity. Moreover, while powerful 
in the abstract, it is inherently unphysical to propagate something 
both classically and quantally. Some of these problems can be avoided 
by monitoring the quantum entropy, which does not measure distances 
but directly addresses relevant issues of information. The linear or 
Renyi entropy of second order $S_2$ is also the natural logarithm of the 
purity $P$ as $S_2 = \ln(P)= \ln[ Tr\{\hat\rho^2\}]$. Note that
$P={2\pi\hbar}Tr\{\rho_W^2\}$ where the ${Tr}$ now represents integration 
over all phase-space variables.  This has been extensively studied and 
for a system with a classically 
chaotic limit, it has been argued\cite{zp,MP} that in the weak-noise, 
small-$\hbar$ classical limit, $-dS_2/dt$ equals the sum of the 
positive classical Lyapunov exponents.  More careful considerations 
generalize this to a weighted sum over Lyapunov exponents\cite{akp}. 
For the classical limit itself, this should arguably be further 
generalized to time-dependent versions\cite{Vulpiani05}. 
That is, although the previous results apply in some limits or special 
cases, even the classical behavior is not fully understood. Less is known 
about the quantum system, particularly the impact of changing scale or
noise through $\hbar, D$, which is what we address below. We work with the 
Hamiltonian $H = \frac{p^2}{2m} -Bx^2 + \frac{C}{2}x^4+ Ax\cos(\omega t)$. 
This is the Duffing oscillator, which as a 1-dimensional driven problem
with a quartic nonlinearity is one of the simplest flows with a rich
phase-space structure and hence is a paradigmatic problems in Hamiltonian 
chaos. The quantum version has also been frequently studied, including 
for decoherence issues\cite{MP,Shiz}. 
\begin{figure}[htbp]
\centerline{\includegraphics[width=9.5cm,height=7.5cm,clip]{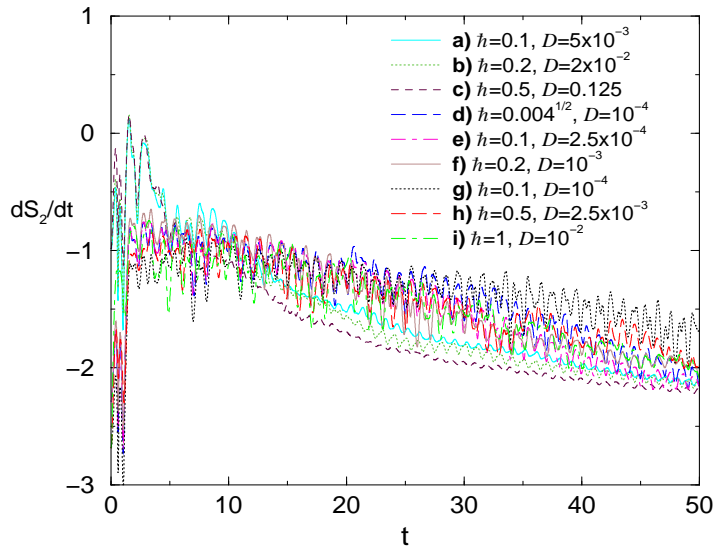}}
\caption{Entropy production rate $\dot{S_2}$ (where $S_2=\ln(P)$ for 
states with purity $P$) for the quantum Duffing oscillator with 
$m=1,B=10,C=1,A=1,\omega=5.35$ and $\hbar,D$ as indicated, showing 
a wide variation in behavior. The 
initial conditions are Gaussians in the chaotic region with 
$\langle x\rangle_{t=0}=1.0$, the spread $\sigma^2_x =0.05, 
\langle p\rangle_{t=0}=0.0$ and the spread $\sigma^2_p$ is set by 
the constraint of defining a minimum-uncertainty state and hence by 
the particular value of $\hbar$. 
}
\label{figone} 
\end{figure}
We briefly review the behavior of the classical density $\rho_c$ in the 
limit of only the $L_c +T$ evolution.  As a result of chaos due to $L_c$ alone, $\rho_c$ increases fine-scale structure exponentially rapidly, 
with a rate given by a generalized Lyapunov exponent. When the structure 
gets to sufficiently fine scales, the noise $T$ becomes important, and it 
acts to decrease, or coarse-grain, fine-scale structure. These 
competing effects can be profitably studied using the measure 
\begin{equation}
\chi^2 \equiv -\frac{{\rm Tr}[\rho_c \nabla^2\rho_c]}{{\rm Tr}
[(\rho_c)^2]} = \frac{{\rm Tr}[|\nabla \rho_c|^2]}{{\rm Tr}
[(\rho_c)^2]}
\label{chi-def}
\end{equation}
whence $\chi^2$ is approximately the mean-square radius of the Fourier 
transform of $\rho$, measuring the structure in the distribution~\cite{97_1}. 
Most importantly, Eq.~(\ref{wigner}) yields the identity 
$dS_2/dt = -2D\chi^2$\cite{footnote-nabla,MP} with this valid
classically or quantum mechanically, that is, with both
$S_2,\chi^2$ computed for $\rho_c$ or $\rho_W$\cite{akp} respectively. 
For a classically uniformly chaotic system, the dynamics of $\chi^2$ can 
be written approximately~\cite{physicaD} as a competition between chaos 
and diffusion as
\begin{equation}
\label{chi-dyn}
\frac{d \chi^2}{dt} \approx 2\Lambda\chi^2 - 4D\chi^4.
\end{equation}
This implies that $\chi^2$ settles after a transient to the 
metastable (that is, constant for finite-time) value 
\begin{equation}
\chi^{2*} = \Lambda/2D
\label{meta}
\end{equation}
where $\Lambda$ is a $\rho$ dependent generalized Lyapunov 
exponent~\cite{schlogl,physicaD}. This {\em classical} argument 
leads to the argument\cite{zp,akp,MP} that quantum entropy-production 
rates are equal to generalized Lyapunov exponents. This applies to a 
greater range of parameters than might be anticipated because 
decoherence suppresses quantum effects. 
While this behavior has been shown in several instances, it does not 
capture the complete picture, particularly the effect of changing $\hbar,D$.
We show this in Fig.~(\ref{figone}) plotting $dS_2/dt$ for the Duffing 
problem with $m=1,B=10, C=1, A=1,\omega = 5.35$, as previously used\cite{MP}. 
The behavior, over a wide parameter and time range is quite complicated.
If a subset (all of those with $\hbar=0.1$) are plotted for a short 
time ($t<15$) as in\cite{MP}, they show the classical Lyapunov exponent 
entropy-production behavior\cite{zp,akp,MP}. This is valid only for some 
small range of parameters and short times. There has been a suggestion
of a superposition of classical and quantal exponential decay\cite{petit} 
for the purity. This would lead to a crossover transition within a
fairly narrow range from one constant value to another in Fig.~(1), which 
we do not see. Other ways of considering the data (as in Fig.~(2) below) 
also do not support this. In general the search for these small regimes 
of linear decay for entropy is not as helpful as understanding the broader 
parameter dependence. 

To do this, consider as in Fig.~(2), $Tr\{\rho_W^2(t)\}/Tr\{\rho_W^2(0)\}$. 
Since the $y$ axis is logarithmic, we are effectively looking at 
$\ln(Tr\{\rho_W^2(t)\})- \ln(Tr\{\rho_W^2(0)\}) = S_2(t) - S_2(0) = S(t)$ 
for our pure-state Gaussians. This shows useful organization invisible 
in Fig.~(1) due to the small-scale variation in a narrow range. Most 
interestingly, the entropy dynamics for the wide variety of parameters 
considered is captured entirely for the times shown by the composite 
parameter $\hbar^2/D\equiv\zeta_0$, even though a wide range of behavior, 
not obviously characterized as exponential decay, is seen as $\zeta_0$ 
is varied. Larger $\zeta_0$ corresponds to high $\hbar$ or low 
noise $D$ or both, and remains closer to a pure quantum state for 
longer times, which makes physical sense. Note also that there is 
some $\zeta_0$ dependence for the time-scale of scaling, with a 
long-term separation of curves.

\begin{figure}[ht]
\centerline{\includegraphics[width=9.5cm,height=7.5cm,clip]{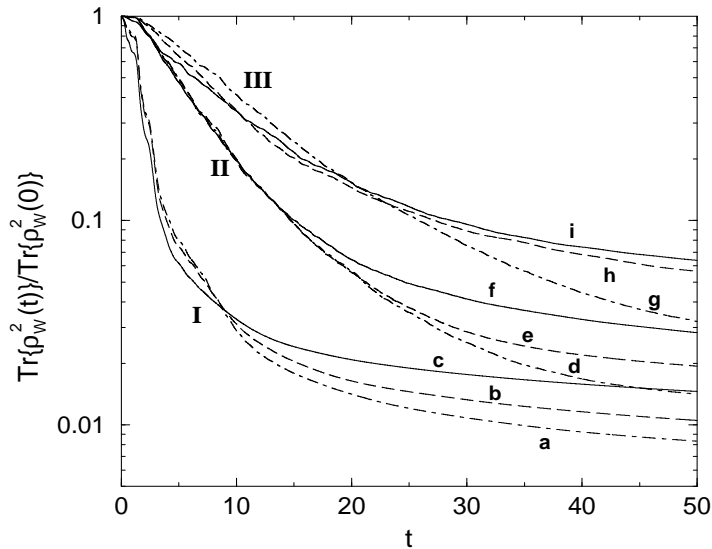}}
\caption{Evolution of the normalized purity for the same states as Fig.~(1) 
in the quantum Duffing oscillator. Scaling is observed relative to the 
parameter $\zeta_0\equiv\hbar^2/D$. 
(I)'Classical': $\zeta_0=2$
a)$\hbar=0.1, D=5\times10^{-4}$ 
b)$\hbar=0.2, D=2\times10^{-2}$,
c)$\hbar=0.5, D=0.125$, 
(II)'Semi-classical': $\zeta_0=40$ 
d)$\hbar=0.004^{1/2}, D=10^{-4}$,
e)$\hbar=0.1, D=2.5\times 10^{-4}$, 
f)$\hbar=0.2, D=10^{-3}$,
(III)'Quantum': $\zeta_0=100$ 
g)$\hbar=0.1, D=10^{-4}$, 
h)$\hbar=0.5, D=2.5\times 10^{-3}$, 
i)$\hbar=1, D=10^{-2}$. 
} 
\label{figtwo} 
\end{figure}
We understand this $\zeta_0$ dependence by considering quantum 
corrections to the classical dynamics, which depend 
(see Eq.~(\ref{wigner})) on the derivatives of the Wigner function. 
Given that the second derivatives $\partial^2\rho_W \propto \chi^2$, 
these corrections scale as 
\begin{equation}
L_{q} \approx \hbar^{2n} \frac{\partial^{2n+1} V(q)}{\partial q^{2n+1}}\;
\frac{\partial^{2n+1} \rho_W}{\partial p^{2n+1}} \approx
\hbar^{2n}\chi^{2n+1}V^{(2n+1)}(x) ,
\label{quant-correction}
\end{equation} 
where $V^{(r)}$ denotes the $r$th derivative of $V$. When the
phase-space distribution hits a metastable state such that $\chi^2$ 
settles to the fixed value $\Lambda/2D$, the difference between 
the quantum and classical evolution may be estimated to depend on 
\begin{equation}
\zeta\equiv\hbar^{2n} \Lambda^{n+1/2}D^{-(n+1/2)} V^{(2n+1)}(x)
\label{zeta-estimate}
\end{equation} 
where, since $\chi$ is a 'length' in Fourier space, we have 
that $x\approx \chi^{-1}=\sqrt{2D/\Lambda}$. This is essentially the
same result as that derived in Ref.~\cite{Shiz} from a completely
different perspective and is also the root of the suggestion in
Ref.~\cite{akp-03} to search for scaling. Therefore, the first order
quantum corrections in a semi-classical regime should scale, in 
complete generality, with the single parameter $\zeta$. The particular 
form of $\zeta$ is decided by the details of the Hamiltonian and the 
difference between the quantal and classical propagators. For the 
Duffing problem, the only quantum term of Eq.~(\ref{quant-correction}) 
comes from the $3$rd derivative of the quartic term whence 
Eq.~(\ref{zeta-estimate}) gives that the quantum term goes 
as $\zeta = \hbar^2\chi^2$; for any other form of the potential, 
we expect different corrections and hence different scaling as below.  

We now use this in an expansion technique for entropy dynamics that may 
be applied in general. In the Duffing problem, even though $\dot S_2$ 
is not a simple function of $\zeta$, a scaling relationship still 
obtains in the two parameters $\hbar,D$ as follows. To zeroth order, 
the classical and quantal phase-space distributions are the same, 
$\rho_{W0} \approx \rho_c$, and $\chi_{q0}^2 = \chi_c^2$, where the 
entropy production rate $\dot S_{2q0} = -2 D \chi_{q0}^2$ and the 
numerical subscripts on $\chi_q, \rho_W, \dot S_{2q}$ indicates the 
order of the approximation. We now use the results from 
Eq.~(\ref{quant-correction},\ref{zeta-estimate}) that the 
quantum-classical distance for this system behaves as $\hbar^2\chi^2$. 
To first order we insert the zeroth order solution in this to write 
\begin{equation}
\label{correction}
\rho_{W1} \approx \rho_{W0} + a\hbar^2\chi^2\rho_{W0} = \rho_{c} +
a\hbar^2\chi^2\rho_{c} 
\end{equation} 
where $a$ is constant for the meta-stable state, but time-dependent in
general. We substitute this in Eq.~(\ref{chi-def}) to get that 
$\chi_{q1}^2 \approx \chi_c^2 + a\hbar^2\chi_c^4.$
Corrections from the denominator of Eq.~(\ref{chi-def}) are of 
higher order, and also tend to cancel the higher order corrections 
from the numerator. We insert this first order quantally corrected form 
for the dynamical term into Eq.~(\ref{chi-dyn}) to get that to first 
order in $\hbar^2$, $\chi^2$ obeys 
\begin{equation}
\label{q1-chi-dyn}
\frac{d \chi^2}{dt} \approx 2\Lambda(\chi^2 + a\hbar^2\chi^4) - 4D\chi^4
\end{equation}
and in parallel to Eq.~(\ref{meta}) we get that 
\begin{equation}
\label{q1-chi}
\chi^{2*} = \frac{\Lambda}{2D(1 - \frac{a\hbar^2}{4D})}
\end{equation}
leading finally to 
$\dot S_{2q1} = -2D\chi^2_{q1} = -\Lambda(1 + \frac{a\hbar^2}{4D})$
that is, the quantum correction scales as $\hbar^2/D$.  This expansion 
around the metastable state can occur only when the growth of structure 
is balanced by noise, only when $\chi^2$ is large enough that the 
diffusion term becomes relevant. Since $a$ is in general time-dependent, 
at each value of $\zeta_0$ we expect a different entropy dynamics, as 
in fact we see. In sum, this expansion for the entropy 
dynamics around the metastable state yields a $\hbar^2/D$ dependence for 
entropy, although the time-dependence itself is not easy to extract.

This expansion must fail for arbitrarily large $\zeta_0$, in the quantum 
regime. Here an alternate approach applies: the Poisson bracket term is 
neglected and the dynamics are given approximately by the competition 
between the $L_q$ and $T$ terms alone. To compare them, consider $L_q$: 
For the Duffing system, there is a third-derivative of $\rho_W$ multiplied 
by $x$ (resulting in this acting like a $2$nd derivative overall), compared 
to the $2$nd derivative from the diffusion term\cite{footnote-III}. This 
means that the terms have essentially the same scale, with quantum dynamics 
continuing to add structure and the noise smoothing it out. The 
entropy-production then depends only on the ratio of the parameters multiplying 
these terms which is again $\hbar^2/D\equiv \zeta_0$. 
This last parameter regime is consistent with recent 
results\cite{petit, Prosen} et al. Finally, consider some details of 
the time-dependence: The rate of purity decay decreases with $\zeta_0$. 
Physically, the time-asymptotic dynamics are dominated by essentially 
classical diffusive behavior, with a common final state (the natural 
invariant measure) for all $\rho_W$. 
Since $Tr\{\rho_W^2(0)\}\propto \hbar^{-1}$ (see above), the time-asymptotic 
value of $Tr\{\rho_W^2(t)\}/Tr\{\rho_W^2(0)\} \propto \hbar^{-1}$. 
With the different rates of purity decay, the system approaches
the time-asymptotic state later as $\zeta_0$ increases. Further, within 
each $\zeta_0$, the different values of $\hbar$ separate out from the 
scaling curve as the final diffusive regime kicks in, as seen in Fig.~(2).

The values of $\zeta_0$ where these regimes change is in general determined 
by the parameters of the potential, i.e. by the quantity labeled as $a$ in 
Eq.~(\ref{correction}) above. Given the continuous behavior as a function 
of $\zeta_0$ the actual transition is subjective. In Fig.~(2) we label
what corresponds to rapidly decohering and hence essentially classical
behavior as (I), the relatively slowly decohering and hence deep quantum
behavior as (III) and in-between `semi-classical' behavior as (II) in the 
three sets of curves with $(\zeta_0 = 2, 100, 40)$ respectively. 
That is, for this potential, empirically $\zeta_0 = \zeta_c \approx 10$ 
sets the approximate upper limit of the rapidly decohering regime (I), and 
by extension the quantum regime (III) kicks in at 
$\zeta_0 =\zeta_q \approx \zeta^2_c \approx 100$.  
We note the same scaling also holds (results not shown) for other 
diagnostics as well as for very different parameters for the Duffing 
oscillator, $A=10,\omega=6.07$, a regime of significantly increased 
chaos\cite{MP}. 

In conclusion, our results strengthen the argument that it is 
valuable to study the behavior of nonlinear open quantum systems 
through the scaling behavior of appropriate diagnostics, as recently 
suggested\cite{akp-03}.  In particular, this is used to study the 
non-equilibrium statistical mechanics of an open quantum system with a 
classically chaotic counterpart over a wide parameter range in $\hbar,D$. 
We show that the entropy dynamics of this system can be dramatically 
different from the broadly-accepted Lyapunov exponent dependence which 
is only valid in the classical limit (and is itself arguably 
suspect\cite{Vulpiani05}). We show a $\hbar^2/D$ scaling in the 
time-dependent entropy dynamics, although the particular form of the 
scaling is expected to depend on the form of the nonlinearity 
in general. 

{\em Acknowledgements}: A.G. is partially supported by FAPESP (Brazil) and 
CNPq (Brazil). A.K.P. acknowledges a CCSA Award from Research Corporation, 
the SIT, Wallin, and Class of 1949 Fellowships from Carleton, and 
hospitality from CiC (Cuernavaca) during this work.


\begin{thebibliography}{99}
\bibitem{footnote-nabla}Our argument assumes environmental coupling 
symmetrically to all phase-space variables, with $T$ of the form 
$\nabla^2$ rather than $\partial_p^2$. This simplifies calculations, 
but note that for a chaotic system there is no difference between the 
behavior of these two forms.
\bibitem{footnote-III} Again, this should be different for different
potentials and result in different scaling functions in this regime.
\end{thebibliography}
\end{document}